# Phase-resolved frequency-domain analysis of the photoemission spectra for photoexcited 1$T$-TaS$_2$ in the Mott insulating charge density wave state

Qianhui Ren[1], Takeshi Suzuki[1,*], Teruto Kanai[1], Jiro Itatani[1], Shik Shin[2,3], Kozo Okazaki[1,3,4 †]

[1]*Institute for Solid State Physics (ISSP), University of Tokyo, Kashiwa, Chiba 277-8581, Japan*
[2]*Office of University Professor, The University of Tokyo, Kashiwa, Chiba 277-8581, Japan*
[3]*Material Innovation Research Center, The University of Tokyo, Kashiwa, Chiba 277-8561, Japan*
[4]*Trans-scale Quantum Science Institute, The University of Tokyo, Bunkyo-ku, Tokyo 113-0033, Japan*

We investigate the nonequilibrium electronic structure of 1$T$-TaS$_2$ by time- and angle-resolved photoemission spectroscopy. We observe that strong photoexcitation induces the collapse of the Mott gap, leading to the photo-induced metallic phase. It is also found that the oscillation of photoemission intensity occurs as a result of the excitations of coherent phonons corresponding to the amplitude mode of the charge density wave (CDW). To study the dynamical change of the band dispersions modulated by the CDW amplitude mode, we perform analyses by using frequency-domain angle-resolved photoemission spectroscopy (FDARPES). We find that two different peak structures exhibit anti-phase oscillation with respect to each other by retrieving the amplitude and phase parts of the FDARPES spectra. They are attributed to the minimum and maximum band positions in energy, where the single band is oscillating between them synchronizing with the CDW amplitude mode. We further find that the flat band constructed as a result of CDW band folding survives during the oscillation while the Mott gap is significantly reduced. Our results suggest the CDW phase is robust, and the lattice modulation corresponding to the CDW amplitude mode dynamically modulates the Mott gap.

Transition-metal dichalcogenide (TMD) is a large class of widely studied layered two-dimensional materials. Many of TMDs show a metal-insulator transition (MIT) and some kind of an ordered ground state. Among them, 1$T$-TaS$_2$ is a prototypical example and shows successive charge-density-wave (CDW) phase transitions [1] [2]. As temperature decreases, it undergoes incommensurate CDW (ICCDW), nearly-commensurate CDW (NCCDW), and commensurate CDW (CCDW) phases below 550, 350, and 180 K, respectively. In the CCDW phase, the 13 Ta atoms are modulated to form a Star of David cluster as schematically shown in Fig. 1(a). The band evolution around the Γ point revealed by angle-resolved photoemission spectroscopy is schematically shown in Figs. 1(b)-1(d) [3] [4] [5]. In the ICCDW phase, an electron band around the M point of the Brillouin-zone corner, which is derived from the Ta 5$d$ orbital, crosses the Fermi level ($E_F$) near the Γ point. In the NCCDW phase, the band is folded to the Γ point due to the CDW potential, and CDW hybridization gaps open at the band-crossing points due to the band folding. As a result, a flat band emerges around $E_F$, which can be regarded as evidence of the CDW band folding. Finally, in the CCDW phase, a band gap opens in the flat band, which has been considered a Mott gap [5].

Furthermore, 1$T$-TaS$_2$ shows superconductivity by

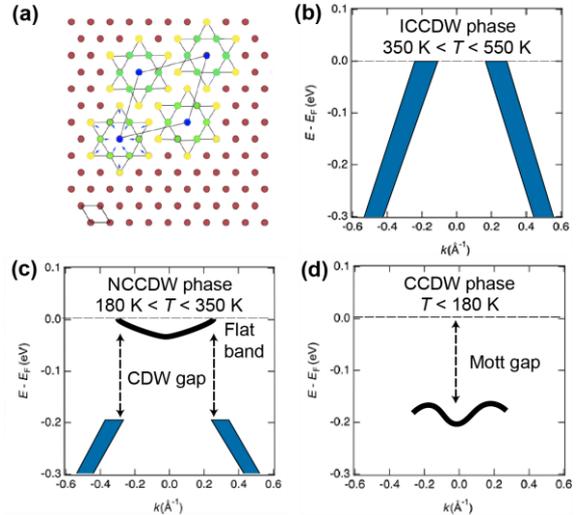

Fig. 1 (a) Schematic illustration for the in-plane lattice modulations in the commensurate CDW phase (CCDW). (b)-(d) Schematic illustration of the band evolution crossing the different phases [5].

applying pressure [6] or chemical doping [7]. These intriguing features such as the coexistence of CDW and Mott-insulating phases or the emergence of superconductivity have motivated intensive studies of 1$T$-TaS$_2$ for understanding the roles of electron-electron and electron-phonon interactions for a long time.

While many experimental methods have been

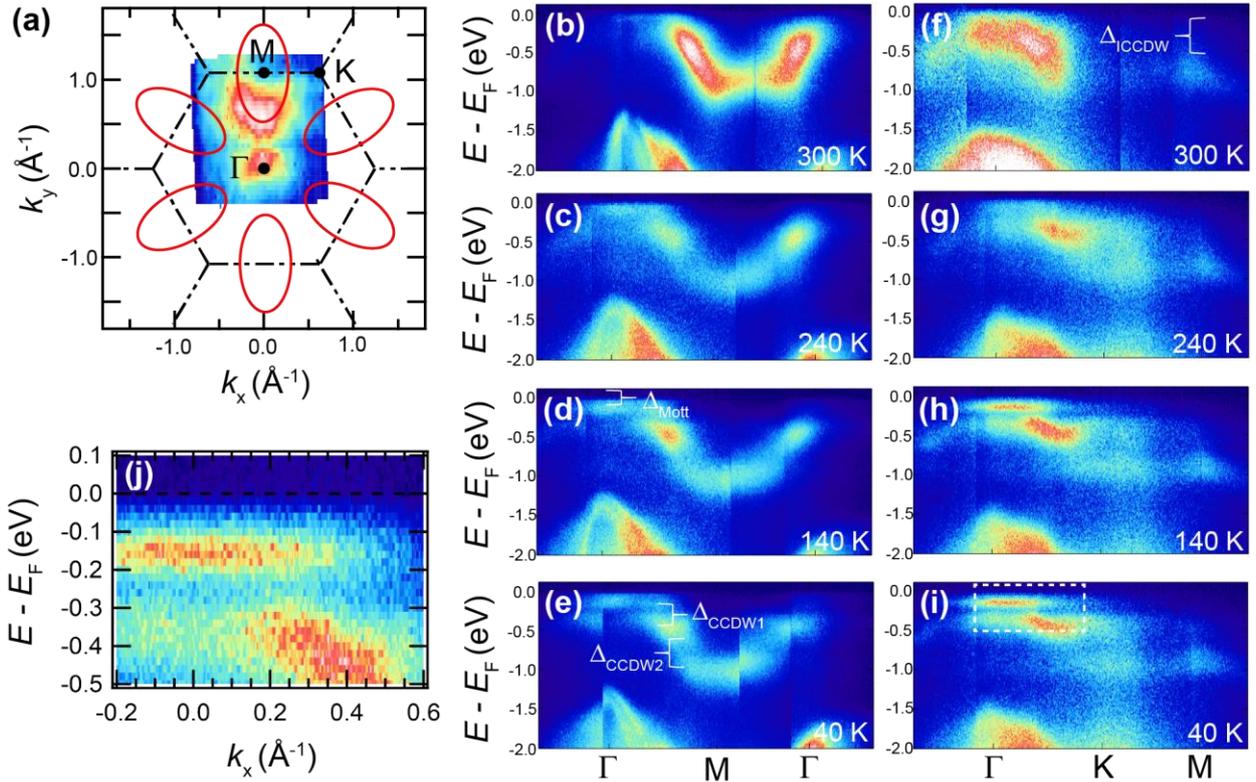

Fig. 2 (a) Fermi surface mapping for 1T-TaS$_2$ taken with a He discharge lamp at 240 K. The FS around the M point is indicated as a red oval. (b)-(i) Temperature-dependent ARPES images taken along the (b)-(e) Γ-M-Γ and (f)-(i) Γ-K-M directions. (j) Enlarged ARPES image corresponding to the area indicated as white box region in Fig. 2(i).

employed to investigate physical properties in equilibrium conditions for 1T-TaS$_2$ [4], ultrafast spectroscopies have provided an alternative route to investigate the nature underlying the above phases via a non-equilibrium state [8]. Furthermore, strong photo-excitation can be found to alter this system to a photo-induced metastable metallic phase at least just after the pump, which suggested the possibility of optoelectronic devices and triggered intensive works [9] [10] [11]. Apart from the quest for such a hidden state, many ultrafast methods have been performed to investigate the interplays between electrons and phonons by perturbing or melting the charge order and analyzing the temporal response [12]. In this regard, time- and angle-resolved photoemission spectroscopy (TARPES) is a very powerful method because it can track the temporal band structure and gain direct information on nonequilibrium electronic states [13] [14] [15]. Especially in 1T-TaS$_2$, where the Fermi surface (FS) is located around the M point in the high-temperature metallic phase, TARPES with the large photon energy obtained by high harmonic generation (HHG) from noble gas to reach the entire 1st Brillouin zone is very ideal [16] [17] [18] [19].

While many previous works using HHG TARPES

revealed the striking and fundamental electronic properties in 1T-TaS$_2$ including the initial melting dynamics of the Mott and CDW phases associated with the collapse of the Mott and CDW gaps [18], the band dispersions dynamically affected by the CDW amplitude mode, and their relations to the Mott and CDW phases are still missing. The reason might lie in the difficulty distinguishing them in time-domain spectra. Recently we developed ansuu analysis method, which we call frequency-domain ARPES (FDARPES) [20]. This method allows us to distinguish the coexisting bands by selectively detecting band dispersions in the frequency domain during the photoinduced phase transition.

In this work, we applied TARPES and FDARPES methods to 1T-TaS$_2$ to investigate the dynamical properties of the Mott and CDW phases associated with the Mott gap and the flat band structure. We directly observe photo-induced melting of the Mott gap in the momentum space and the oscillating behavior as a result of the A$_{1g}$ coherent-phonon excitations in the TARPES image. Strikingly, further FDARPES analysis allows us to find out that the flat band structure survives during the oscillation due to the CDW amplitude mode reducing the Mott gap. Besides, the phase of FDARPES revealed the antiphase behavior between different two peaks, clearly labeling the temporal order of these peaks.

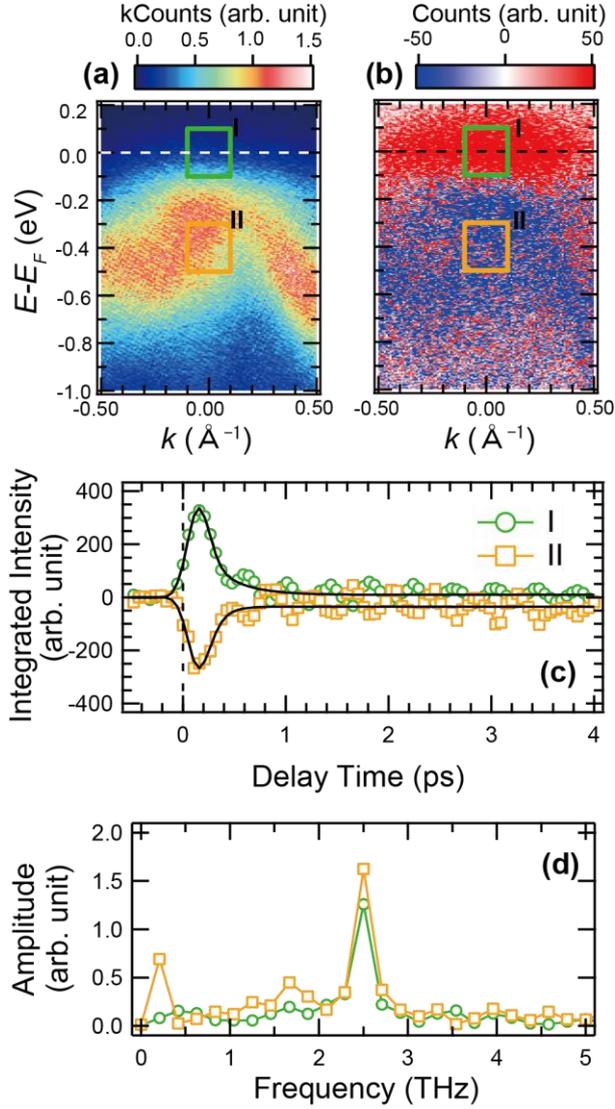

Fig. 3: (a) TARPES image before the arrival of the pump pulse. (b) Differential TARPES image at the delay time of 240 fs. (c) Time-dependent photoemission intensity integrated at the energy-momentum region denoted as I and II in (a) and (b). (d) Amplitudes of the Fourier transform of the oscillatory components deduced from I and II in (c).

For the static ARPES, we used a photon energy of 21.2 eV obtained from a He discharge lamp (Scienta Omicron VUV 5050) and detected the photoelectrons using a Scienta R4000 analyzer. In the TARPES measurements, the experimental setup is similar to that of our previous reports [21] [22] but updated to the 10 kHz laser system(Spectra Physics, Solstice Ace) [23] [24]. The pulse duration of the laser is 35 fs, the photon energy of the probe light is 21.7 eV corresponding to the 7th order high harmonic generation, the total energy resolution is set to 250 meV, and the temporal resolution is evaluated to be ~70 fs. Details of the setup are described elsewhere [25]. All the spectra for the TARPES measurements are taken at the temperature of 15 K. High-quality single crystals of $1T$-TaS$_2$ were provided by HG Graphene. Clean surfaces were obtained by cleaving *in situ*.

To confirm the cleaved surfaces of the sample and temperature-induced phase transitions, we perform the static ARPES measurements. Figure 2(a) shows the Fermi surface (FS) mapping of $1T$-TaS$_2$ measured at 240 K. The FS is clearly observed around the M point, which is indicated as a red oval. The photoemission intensity around the $\Gamma$ point is ascribed to the flat band associated with the NCCDW phase [5]. Figures 2(b)-2(i) show the temperature-dependent ARPES images along the (b)-(e) $\Gamma$-M-$\Gamma$ and (f)-(i) $\Gamma$-K-M directions. With decreasing temperature to 240 K, the flat band structure is more pronounced at the $\Gamma$ point as a result of the band folding signifying the CDW phase [5]. With further decreasing temperature to 140 K, a small gap of the flat band with respect to $E_F$ is noticed, which corresponds to the Mott gap denoted as $\Delta_{Mott}$ [4]. Regarding the CDW phases, the ICCDW gap denoted as $\Delta_{ICCDW}$ around the M point increases with decreasing temperature. Besides, the two gaps associated with the CCDW phase denoted as $\Delta_{CCDW1/2}$ are observed at the band dispersion around the $\Gamma$ point. The Mott and CDW gaps are more clearly seen in the enlarged ARPES image shown in Fig. 2(j), which corresponds to the white-dashed box region indicated in Fig. 2(i). These findings are consistent with the previous report [4].

Now we discuss the TARPES results. Figure 3(a) shows the TARPES image before the arrival of the pump pulse around the $\Gamma$ point along the M-$\Gamma$-M direction. While the flat band is clearly observed, the CCDW gap at the band crossing point due to the CDW folding is not so clear due to the lack of energy resolution, which is typical for the HHG TARPES measurements. Just after the strong pump pulse excitation of 4.2 mJ/cm$^2$, the system turns into the metallic phase [18]. We also confirm that the excited state comes back to the ground state before the arrival of the next pump pulse by comparing TARPES data at the negative delay and static ARPES data shown in Fig. S1 [26], where no metallic behavior in TARPES data is confirmed. Figure 3(b) shows the differential TARPES image at the delay time ($\Delta t$) of 240 fs. The increase and decrease of photoemission intensity are indicated as red and blue colors, respectively. It is noticed that the band dispersion crosses $E_F$ as the red region around $E_F$ increases. This demonstrates the photo-induced IMT with a collapse of the Mott gap as previously reported [18]. The complete set of TARPES images is shown in Fig. S2 [26]. The transient electron temperature at 160 fs is estimated to be higher than 600 K as shown in Fig. S3 [26], which is above the NCCDW transition temperature (350 K), above which the flat band disappears in equilibrium.

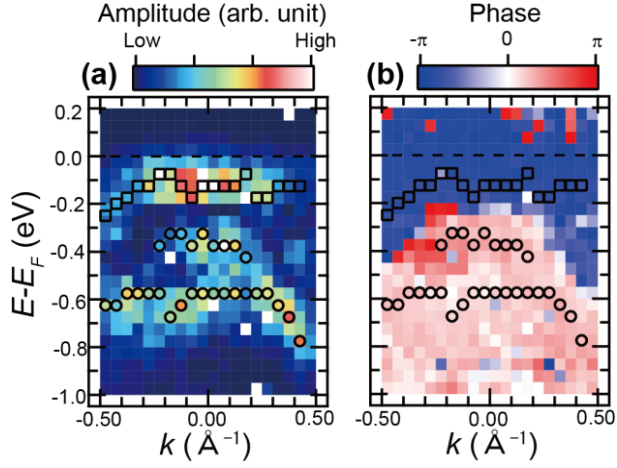

Fig. 4: Frequency-domain ARPES images for (a) amplitude and (b) phase at the frequency of 2.5 THz. Black markers show the peak positions for the amplitude (a).

To study the dynamics in more detail, Fig. 3(c) shows the time-dependent photoemission intensity integrated at the energy-momentum region denoted as I and II in Figs. 3(a) and 3(b). Overall, the immediate increase (decrease) at $\Delta t = 0$ is followed by the fast decrease (increase) and slow relaxation is observed for the region I (II). This behavior is generally explained by the two-temperature model [27]. Besides, significant oscillations are confirmed to be superimposed onto both data. Moreover, these oscillations have an anti-phase character concerning each other. For highlighting the oscillatory components, we first subtract overall dynamics by fitting them to a double-exponential function convoluted with a Gaussian, shown as the black solid lines in Fig. 3(c). Fourier transformations are performed for the subtracted data, and amplitudes for each frequency component are shown in Fig. 3(d). One can see the strong single peak at 2.5 THz, which corresponds to the breathing $A_{1g}$ mode and is also called the CDW amplitude mode, confirmed by Raman spectroscopy [28], and the oscillation is as a result of coherent-phonon excitations based on the displacive excitation mechanism [29]. It should also be mentioned that the oscillation revealed t by the TARPES measurements is due to the electron-phonon coupling, which connects the modulations between the lattice structure and electronic wave functions [30].

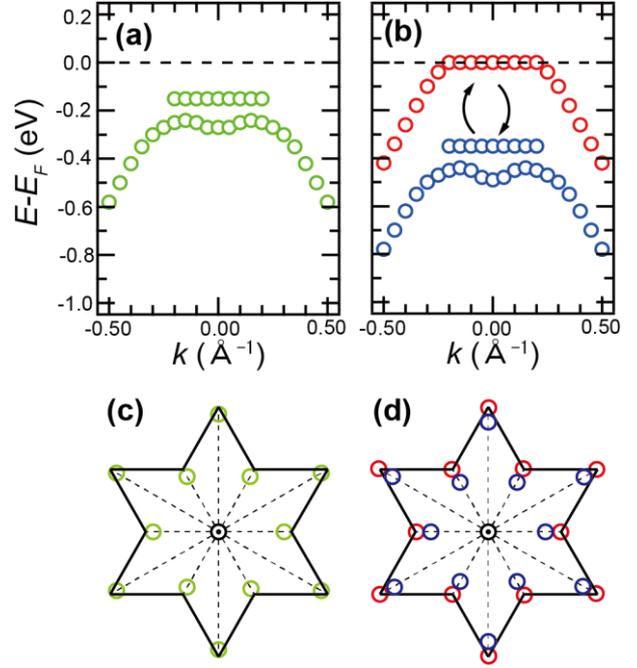

Fig. 5: Schematic illustration of the band dispersions at the (a) equilibrium and (b) after photoexcitation. Schematic illustration of the displacement of Ta atoms at the (c) equilibrium and (d) after photoexcitation. Red and blue markers in the band dispersions shown in (b) corresponds to the displacement of Ta atoms shown as red and blue markers in (d).

To further investigate the dynamical change of the electronic band dispersions by the CDW amplitude mode, we proceed to FDARPES analysis. Figures 4(a) and 4(b) show the FDARPES images for amplitude and phase, respectively, at the frequency of 2.5 THz around the $\Gamma$ point. FDARPES images at the different momentum cuts are shown in Fig. S4 [21]. One can notice that there exist two distinctive peak structures near and below $E_F$ shown as black squares and circles, respectively, in Fig. 4(a). Interestingly, their phases shown in Fig. 4(b) are different by nearly $\pi$, which was already seen as an anti-phase behavior in the time-dependent photoemission intensity in Fig. 3(c). This anti-phase character clearly shows that the single band oscillates between these two peak structures, synchronizing with the lattice modulation corresponding to the CDW amplitude mode, which is schematically shown in Fig. 5. The separation into the flat and M-shaped bands when the band shifts to the energetically lower position is due to the opening of the CDW gap, which is also seen at the temperature-dependent static ARPES shown in Fig. 2.

Strikingly, the shape of the band drastically changes between these two peak positions, and the flat band structure with a smaller Mott gap is seen near $E_F$ shown as black squares in Fig. 4. As seen in the temperature-dependent static ARPES shown in Fig. 2, the flat band structure is the signature of the CDW phase. Based on this connection, the observed flat band in the FDARPES spectra strongly suggests that the CDW band folding survives even though the Mott phase is strongly suppressed seen as a reduction of the Mott gap. Furthermore, it is unveiled that the CDW amplitude mode can dynamically modulate the Mott gap corresponding to the position of the band dispersions with respect to the $E_F$ as well as the CDW gap corresponding to the distance between the flat and M-shaped band schematically seen in the blue circles in Fig. 5(b). These dynamical change of the band structure during the oscillation of the CDW amplitude mode are captured by FDARPES spectra. It is also of notice that these clear peaks are only elucidated in the frequency domain, never extracted in the time domain analysis as clearly noticed by seeing the TARPES image in Fig. S2. The FDARPES technique has a similarity to lock-in amplification in that it can significantly improve the signal-to-noise ratio by extracting a specific frequency component.

In summary, we have conducted TARPES measurements on 1$T$-TaS$_2$ and proceeded to perform FDARPES analysis to understand the underlying nature of the photo-induced state. While the Mott-insulating gap is closed just after pump confirmed by TARPES, the flat band dispersion is clearly seen in the FDARPES image, which strongly indicates survival of the CDW band folding even during the photo-induced state. Our method can be used in many systems and reveal hidden properties by investigating the frequency domain.

See the supplementary material for Figs. S1, S2, S3, and S4.

This work was supported by Grants-in-Aid for Scientific Research (KAKENHI) (Grant Nos. JP19H01818, JP19H00659, and JP19H00651) from the Japan Society for the Promotion of Science (JSPS), by JSPS KAKENHI on Innovative Areas "Quantum Liquid Crystals" (Grant No. JP19H05826), and the Quantum Leap Flagship Program (Q-LEAP) (Grant No. JPMXS0118068681) from the Ministry of Education, Culture, Sports, Science, and Technology (MEXT), and by research grants from The Murata Science Foundation and The Hattori Hokokai Foundation.

## AUTHOR DECLARATIONS
### Conflict of interest
The authors have no conflicts to disclose.

## Author contributions
**Qianhui Ren**: Conceptualization (lead); Data curation (lead); Formal analysis (lead); Investigation (lead); Methodology (lead); Software (lead); Writing – review & editing (equal). **Takeshi Suzuki**: Formal analysis (equal); Investigation (equal); Methodology (equal); Writing – original draft (lead); Writing – review & editing (lead) Supervision (lead); **Kozo Okazaki**; Conceptualization (lead); Investigation (lead); Methodology (lead); Project administration (lead); Supervision (lead); Writing – review & editing (lead). **Teruto Kanai**; Investigation (equal); Methodology (equal); Writing – review & editing (equal). **Jiro Itatani**; Investigation (equal); Methodology (equal); Writing – review & editing (equal). **Shik Shin**; Conceptualization (lead); Methodology (lead); Project administration (lead); Supervision (lead).

### Data availability
The data supporting the findings of this study are available from the corresponding author upon request.

*Corresponding author.
takeshi.suzuki@issp.u-tokyo.ac.jp
†Corresponding author.
okazaki@issp.u-tokyo.ac.jp